\documentclass[11pt,a4paper]{article}

\usepackage{jheppub}

\usepackage[english]{babel} 
\usepackage{amssymb}
\usepackage{graphicx}
\usepackage{amsmath}
\usepackage{commath}
\usepackage{amssymb}
\usepackage{braket}
\usepackage{ dsfont }
\usepackage{physics}
\usepackage{xcolor}
\usepackage[titletoc, title]{appendix}

\usepackage{tikz}
\usetikzlibrary{positioning}
\usetikzlibrary{intersections}
\usetikzlibrary{fadings} 

\usepackage{subfigure}
\usetikzlibrary{backgrounds,automata}

\newcommand{\lp}{\left(}
\newcommand{\rp}{\right)}

\newcommand{\asinh}{\,\mathrm{arcsinh}}

\newcommand{\tilt}{\tilde{t}}

\title{\boldmath From the BTZ black hole to JT gravity: geometrizing the island}

 \author{Evita Verheijden}
 \author{and Erik Verlinde}
 \affiliation{Institute of Physics \&
Delta Institute for Theoretical Physics,\\ University of Amsterdam, Science Park 904, \\
1090 GL Amsterdam, The Netherlands\\
}

\emailAdd{e.m.h.verheijden@uva.nl}
\emailAdd{e.p.verlinde@uva.nl}

\abstract{We study the evaporation of two-dimensional black holes in JT gravity from a three-dimensional point of view. A partial dimensional reduction of AdS$_3$ in Poincar\'e coordinates leads to an extremal 2D black hole in JT gravity coupled to a `bath': the holographic dual of the remainder of the 3D spacetime. Partially reducing the BTZ black hole gives us the finite temperature version. We compute the entropy of the radiation using geodesics in the three-dimensional spacetime. We then focus on the finite temperature case and describe the dynamics by introducing time-dependence into the parameter controlling the reduction. The energy of the black hole decreases linearly as we slowly move the dividing line between black hole and bath. Through a re-scaling of the BTZ parameters we map this to the more canonical picture of exponential evaporation. Finally, studying the entropy of the radiation over time leads to a geometric representation of the Page curve. The appearance of the island region is explained in a natural and intuitive fashion.}

\begin{document} 
\maketitle

\flushbottom

\section{Introduction}\label{sec:intro}

Recent developments have shed new light on the question whether the formation and evaporation of black holes happen in a unitary fashion. This problem, known as the black hole information problem, boils down to reproducing the Page curve: to save unitarity, the von Neumann entropy of the collected Hawking radiation should initially rise during evaporation, but start decreasing after reaching a halfway point known as the Page time \cite{Page1, Page2}. However, a semiclassical gravity approach gives an ever-increasing entropy, usually called the Hawking curve. This tension has a formal resolution in holography: replacing the black hole by its dual representation, it is clear that the process must be unitary. One would like to find the analogous statement from the gravity point of view, e.g.\ using holographic entropy tools such as the Ryu-Takayanagi formula and its covariant extension \cite{RyuTakayanagi, HubenyRT}. The recent progress was initiated by two groups \cite{Penington1, AlmheiriEMM} that used the minimal Quantum Extremal Surface (QES) \cite{EngelhardtQES}, the surface that minimizes the generalized entropy, to monitor the evolution of the entropy of the evaporating black hole. 

It was found that the QES undergoes a transition at the Page time and jumps from being the empty surface to a surface just inside the black hole horizon. This reproduces a Page curve for the black hole in geometric terms, but not for the radiation. Using a doubly holographic model, i.e.\ considering a black hole in a gravitational theory with matter that itself has a holographic dual, the authors of \cite{AlmheiriMMZ} showed that in this setup one can simply use the RT/HRT formula in the higher-dimensional dual. This ensures that in the evaporating black hole picture the minimal surface of the radiation and the black hole coincide, thus obtaining a Page curve for the radiation as well. The higher-dimensional geometry connects the radiation to the black hole interior, such that (at late times) the black hole interior becomes part of the entanglement wedge of the radiation or `bath'. This led to a new rule for computing the entropy in gravitational systems, now known as the ``Island Rule": 
\begin{equation}
	S[\text{Rad}]^{\rm QG} = \underset{I}{\text{min}}\Big\{ \underset{I}{\text{ext}} \Big[ S[\text{Rad} \cup I]^{\rm SCG} + \frac{\text{Area}[\partial I]}{4G}\Big] \Big\}~.
\end{equation}
The prescription tells us that to compute the entropy of Hawking radiation in quantum gravity, we should include ``quantum extremal islands" in our semi-classical entropy calculation. These islands can minimize the entropy, e.g.\ an island just inside the black hole horizon will include Hawking partners of the radiation. The price to pay is the area of the island. Finally, one has to extremize and minimize over all possible islands. 

Most of the recent work takes place in two spacetime dimensions and in Anti-de Sitter space, which makes the calculation of entanglement entropy more tractable.\footnote{There is no reason to believe the general arguments should fail in higher dimensions, and some explicit extensions to higher-dimensional systems exist, see e.g.\ \cite{higherD1} for a numerical example, also following the doubly-holographic route, and \cite{higherD2} for an exact example. Extensions to black holes in asymptotically flat space also exist, see e.g.\ \cite{flat1, flat2}.}\ For example, in \cite{AlmheiriEMM}, the specific model considered was that of two-dimensional JT gravity coupled to a bath. In the doubly-holographic model of \cite{AlmheiriMMZ}, JT gravity was coupled to conformal matter (a CFT$_2$), which played the role of the bath. This CFT is then itself the boundary theory of a dual AdS$_3$ bulk. The JT gravity lives on a Planck brane, which in the three-dimensional picture can be thought of as a Randall-Sundrum or End-of-the-World (EOW) brane. These doubly-holographic models, and the closely related BCFT models, have proven successful in other setups as well \cite{RaamsdonkBCFT, Janus, BCFT, Easy1, Easy2, Easy3, infoflow, flat2}. 

The natural environment to study the information paradox consists of evaporating black holes. However, a version of the Page curve can also be found for eternal black holes. In \cite{RaamsdonkBCFT}, a transition in the entanglement entropy was shown to occur also for the case of thermal equilibrium between black hole and bath: even though the black hole does not evaporate, it does ``radiate" information away. In a related paper \cite{AlmheiriOutside}, the authors use the language of islands to study a version of the information paradox. Interestingly, they show that the island that resolves the paradox lies \textit{outside} of the horizon.

In this paper we like to add to this exciting line of research by combining some of the previous techniques. We present a model of evaporation of a two-dimensional black hole that is coupled to a heat bath consisting of a thermal CFT. Microscopically we represent the black hole by a one-dimensional quantum mechanical model that lives on the boundary of the CFT. We will choose a quantum mechanical model that has a dual description in terms of JT gravity with an asymptotically AdS$_2$ geometry. Since the quantum mechanical model has a finite temperature (it is in an excited state), its dual is given by a two-dimensional AdS black hole.  

Our goal is to shed light on the origin of the Page curve and the island phenomenon by geometrizing both concepts in this very simple setting.  To simplify things even further, we will implement the original argument of Page: he envisaged the evaporation process by considering the pure state of the total system and by modeling the evaporation as a time evolution of the way that the total pure state is divided up in into a black hole and radiation part. This version of the Page argument is most direct if one thinks of the total state as a given pure state that itself does not evolve in time. To make this concrete, we identify the pure state of the total system with the final state of the radiation system at the end of the evaporation process. In our context the final state describes an excited CFT state, and hence is dual to the full BTZ geometry. 
The evolution of the system is then represented in the Heisenberg picture by evolving the operators in time. Initially all operators are associated with the black hole system, while as time progresses more and more operators become associated with the thermal bath of the radiation. 

This idea can be concretely implemented in our model by viewing the 
2D black hole in JT gravity as the dimensional reduction of part of the 3D BTZ geometry. In this representation, the angular metric component takes on the role of the dilaton. Instead of applying a full dimensional reduction, we will reduce over part of the range of the angular coordinate; thereby, we effectively split the geometry into two parts: a JT black hole, and a 2D CFT dual to the remainder of the 3D geometry. See figures \ref{fig:geod1} and \ref{fig:BTZJT} for a schematic depiction of the setup. The 2D CFT will model a `bath', and we can now compute the entanglement entropy of an interval in the bath system using RT surfaces. 

We can introduce dynamics into the finite temperature JT black hole system by giving time-dependence to the parameter that controls the dimensional reduction. In this way, we can let the black hole `geometrically' evaporate. From the BTZ perspective, we simply move the dividing line between the degrees of freedom that are `in' and `out' of the black hole. From the JT perspective, the mass decreases linearly and the temperature is fixed. Exploiting a map of BTZ parameters discussed in \cite{deBoerBTZ}, we can equivalently view this evaporation as adiabatically decreasing the mass of the BTZ black hole, such that we can consider it to be in thermal equilibrium at each instant of time. On the JT gravity side, this gives us a more standard exponential evaporation, where the temperature depends on time. Finally, computing the entropy of the entire bath system for the `geometric evaporation', we obtain a Page curve for the radiation entropy. \\

\noindent This paper is organized as follows. In section \ref{sec:reduction} we briefly review  black holes in JT gravity, and discuss how to obtain both extremal and non-extremal black holes from AdS$_3$ by dimensional reduction. In section \ref{sec:entropies}, we compute the generalized entropy for intervals in the black hole plus bath systems from a three-dimensional point of view. Then, we will introduce dynamics in section \ref{sec:dynamics}, and allow the non-extremal black hole to slowly evaporate. Finally, we obtain a Page curve for the entropy of the radiation. 
\flushbottom

\section{\texorpdfstring{JT gravity from {AdS$_3$} black holes}{JT gravity from AdS3 black holes}} \label{sec:reduction}
After a brief reminder on extremal and non-extremal black holes in JT gravity, we show how to obtain these black holes from AdS$_3$ by dimensional reduction. We end with some comments on retrieving the boundary action from the dimensional reduction. 

\subsection{Review: black holes in JT gravity} \label{sec:JT}
Jackiw-Teitelboim (JT) gravity is a two-dimensional dilaton gravitational theory (see \cite{Jackiw, Teitelboim} for the original model, and \cite{ AlmheiriPolchinski, HermanBackreaction} for compact reviews). If we require the classical background metric to be given by AdS$_2$ with AdS radius $\ell$, we find the following action: 
\begin{equation}
	S = \frac{1}{16\pi G} \left[\int d^2 x \sqrt{-g} \Phi_0 R \,\, +  \int d^2 x \sqrt{-g}  \Phi \lp R + \frac{2}{\ell^2}\rp \right] + S_\text{matter}~. \label{eq:JTaction} 
\end{equation}
The first term (with $\Phi_0$ a constant) is topological and determines the extremal entropy; after adding the appropriate boundary term it gives the Euler characteristic of the manifold. The second term is the JT term and the equation of motion for $\Phi$ sets $ R = - \frac{2}{\ell^2}$, i.e.\ it forces the spacetime to be asymptotically AdS$_2$. $S_\text{matter}$ is some arbitrary matter system that couples to the metric but not to the dilaton. The general solution for the metric is  
	\begin{equation} \label{eq:JTgeneralmetric}
		ds^2 = - \frac{4 \ell^2~ \dif X^+ \dif X^-}{(X^+-X^-)^2}~,
	\end{equation}
where $X^+ (u)$ and $X^- (v)$ are general monotonic functions of the lightcone coordinates $(u,v)$. The AdS$_2$ boundary is located at $X^+(u)=X^-(v)$. 

The action \eqref{eq:JTaction} admits black hole solutions, dynamically formed by throwing in matter from the boundary. The vacuum equations of motion are solved by the dilaton profile 
	\begin{equation} \label{eq:JTgeneralDil}
		\Phi = \Phi_0 + 2 \Phi_r \frac{1 - \kappa E X^+X^-}{X^+-X^-}~,
	\end{equation}
where $E = M$ is the mass of the black hole and $\Phi_r = \frac{4 \pi G}{\kappa}$ is an integration constant that specifies the asymptotic boundary conditions of the dilaton field (notice that $\Phi$ is dimensionless, but $\Phi_r$ has dimensions of length\footnote{In some references, e.g.\ \cite{SarosiSYK, MaldacenaNAdS}, the action is derived from a four-dimensional parent theory, such that the Newton's constant has dimension (length)$^2$. This gives the dilaton $\Phi$ dimensions of (length)$^2$ and the interpretation of an area. In \eqref{eq:JTaction}, however, we work directly in two dimensions such that $G$ is dimensionless.}). In particular, close to the boundary $g = \frac{\ell^2}{\epsilon^2}$ and $\Phi - \Phi_0 = \frac{\Phi_r}{\epsilon}$, with $\epsilon$ the UV cutoff. Through the substitution 
	\begin{equation}
		X^+(u) = \frac{1}{\sqrt{\kappa E}}\tanh\lp \sqrt{\kappa E} u\rp~, \qquad X^-(v) = \frac{1}{\sqrt{\kappa E}} \tanh \lp \sqrt{\kappa E} v\rp~, 
	\end{equation}
we find that the metric and dilaton are periodic in imaginary time with period 
	\begin{equation} \label{eq:Htemp}
	\beta = \pi / \sqrt{\kappa E} \quad \Rightarrow  T_H = \frac{1}{\pi}\sqrt{\frac{8 \pi G E}{2\Phi_r}}~.
	\end{equation}
In terms of the lightcone coordinates $(u,v)$, the metric and dilaton profile are
	\begin{equation} \label{eq:JTBH}
		ds^2 = - \frac{4\pi^2 \ell^2 }{\beta^2} \frac{\dif u \dif v}{\sinh^2\frac{\pi}{\beta}(u-v)}~, \qquad \Phi = \Phi_0 + \frac{2\pi \Phi_r}{\beta} \frac{1}{\tanh \frac{\pi}{\beta}(u-v)}~.
	\end{equation}
The future and past horizons are at $u=\infty$ and $v = - \infty$. Note that if we take the limit $E \to 0$ we recover the AdS$_2$ geometry in Poincaré coordinates and the corresponding dilaton profile: 
	\begin{equation} \label{eq:AdS2Poincare}
		ds^2 = - \frac{4 \ell^2 ~ \dif X^+ \dif X^-}{(X^+-X^-)^2}~, \qquad \Phi = \Phi_0 + \frac{2\Phi_r}{(X^+-X^-)}~.
	\end{equation}
Since $E \to 0$ sets also the Hawking temperature $T_H \to 0$, we will interpret this solution as the extremal AdS$_2$ black hole. In the following sections, we will see how to obtain these black holes from three dimensional Anti-de Sitter space. 

\subsection{Dimensional reduction from 3D Einstein to JT gravity}
Consider the three-dimensional action
	\begin{equation} \label{eq:3Daction}
		S = \frac{1}{16\pi G^{(3)}}\int d^3 x \sqrt{-g}  (R^{(3)} - 2\Lambda )~,
	\end{equation}
with negative cosmological constant $\Lambda < 0$. Solutions are asymptotically AdS$_3$ with the AdS radius given by $\Lambda = - \frac{1}{\ell_3^2}$, and include the BTZ solution. Suppose that we have a solution for which the metric field is independent of one coordinate, which we will call $\varphi$, and that it can be written as 
	\begin{equation} \label{eq:3Dmetric}
		ds^2 = g_{\mu\nu} \dif x^\mu \dif x^\nu = h_{ij} (x^i) \dif x^i \dif x^j + \phi^2 (x^i) \ell_3^2 \dif \varphi^2~,
	\end{equation}
where the indices $\mu,\nu = 0,1,2$ and $i,j=0,1$. Then the action \eqref{eq:3Daction} reduces to 
	\begin{equation} \label{eq:reducedaction}
		S = \frac{2 \pi \alpha \ell_3}{16\pi G^{(3)}}\int d^2 x \sqrt{-h} \phi (R^{(2)} - 2 \Lambda)~,
	\end{equation}
See also \cite{AchucarroBTZ}. Here, we accounted for a partial reduction controlled by the parameter $\alpha \in (0,1]$ for reasons that will become clear later. We see that we retrieved the JT action \eqref{eq:JTaction} (ignoring the topological piece) for $\Lambda = - \frac{1}{\ell^2}$, the cosmological constant for AdS$_2$ gravity. If we identify $\ell_3 = \ell$ and $G^{(3)} = \ell  G^{(2)}$ then we are led to conclude that the dilaton in \eqref{eq:JTaction} is given by
	\begin{equation}
		\Phi  = 2\pi \alpha \phi~. 
	\end{equation}
Since the three-dimensional Newton's constant has dimensions of length, from this reduction we again inherit a dimensionless dilaton (remember that in \eqref{eq:JTaction}, the two-dimensional Newton's constant is dimensionless, and therefore the dilaton as well). Note that we do not obtain the topological part of the JT action. Since we are interested in the fluctuations away from extremality, this will not pose a problem. Thus, we will define the boundary condition to be 
	\begin{equation}
		\Phi |_{bdy} = \frac{\Phi_r}{\epsilon}~.
	\end{equation}
Taken to the boundary we have
	\begin{equation}
		\phi |_{bdy} = \frac{\ell}{\epsilon}~,
	\end{equation}
such that we should interpret
	\begin{equation}
		\Phi_r = 2\pi \ell \alpha \equiv \Phi_r^0 \alpha~. 
	\end{equation}
In what follows we will therefore use
	\begin{equation}
		\Phi = \Phi_r \frac{\sqrt{g_{\varphi\varphi}}}{\ell^2} = \Phi_r \frac{\phi}{\ell}~.
	\end{equation}
We will now apply this procedure to Poincar\'e AdS$_3$ and the BTZ black hole. 

\subsection{\texorpdfstring{Extremal AdS$_2$ black hole from AdS$_3$}{Extremal AdS2 black hole from AdS3}}
The metric of AdS$_3$ in Poincar\'e coordinates is given by: 
\begin{equation}
	ds^2 = \frac{\ell^2}{z^2}(-\dif t^2 + \dif z^2 + \dif x^2)~.  \label{eq:Poinc}
\end{equation}
We would like to reproduce \eqref{eq:AdS2Poincare}. First, note that we can use the coordinate transformation $z = \frac{\ell^2}{r}$ and $x = \ell\varphi$ to rewrite the above as
	\begin{equation} \label{eq:Poincarecylin}
		ds^2 = - \frac{r^2}{\ell^2} \dif t^2 + \frac{\ell^2}{r^2} \dif r^2 + r^2 \dif \varphi^2~. 
	\end{equation}
This is exactly of the form \eqref{eq:3Dmetric}, with $\phi \ell = r$. Hence we arrive immediately at the conclusion that AdS$_3$ in Poincar\'e coordinates reduces to a solution of JT gravity. To get exactly \eqref{eq:AdS2Poincare}, we can instead use lightcone coordinates $X^\pm = t \pm z$, in which the metric \eqref{eq:Poinc} becomes
	\begin{equation} \label{eq:PoincLCcoord}
		ds^2 = \frac{-4 \ell^2 \,\dif X^+ \dif X^-}{(X^+-X^-)^2} + \frac{4\ell^4 \,\dif\varphi^2}{(X^+-X^-)^2}~.
	\end{equation}
Therefore, comparing to \eqref{eq:AdS2Poincare} we obtain precisely the AdS$_2$ Poincar\'e metric if we identify the dilaton:
	\begin{equation} \label{eq:PoincareDil}
		\Phi = \Phi_r \frac{\sqrt{g_{\varphi\varphi}}}{\ell^2} =  \frac{2\Phi_r}{X^+-X^-}~.
	\end{equation}
Note that this matches the dilaton in \cite{AlmheiriOutside} up to the extremal part of the dilaton $\Phi_0$. 

\subsection{Finite temperature \texorpdfstring{AdS$_2$}{AdS2} black hole from BTZ}
We can follow a similar procedure for the finite temperature case. Now, we start from the BTZ geometry, 
	\begin{equation}
		ds^2 = - \lp \frac{r^2-R^2}{\ell^2} \rp \dif t^2 + \lp \frac{r^2-R^2}{\ell^2} \rp^{-1} \dif r^2 + r^2 \dif \varphi^2~, \label{eq:BTZ}
	\end{equation}
where $R^2 = 8 G M \ell ^2$ is the horizon radius and the inverse temperature is $\beta = \frac{2\pi \ell^2}{R}$. The BTZ metric \eqref{eq:BTZ} is also of the form \eqref{eq:3Dmetric} and hence we could immediately identify again $\phi \ell = r$. To make contact with our earlier description of the non-extremal black hole in JT gravity, we change coordinates to $u = t + r^*$, $v=t-r^*$. Here $r^*$ is the usual tortoise coordinate, defined through
	\begin{equation}
		\dd r^{*} = \lp \frac{r^2-R^2}{\ell^2} \rp^{-1} \dd r~. 
	\end{equation}
Outside the horizon, the metric takes the form 
	\begin{equation} \label{eq:BTZmetric2}
		ds^2 = - \frac{4\pi^2 \ell^2}{\beta^2} \frac{\dif u \dif v}{\sinh^2 \frac{\pi}{\beta}(u-v)} + \frac{4\pi^2 \ell^4}{\beta^2} \frac{1}{\tanh^2{ \frac{\pi}{\beta}(u-v)}} \dif\varphi^2~.
	\end{equation}
In the first part we recognize precisely the AdS$_2$ black hole metric \eqref{eq:JTBH}. Furthermore we can identify again the dilaton profile
	\begin{equation} \label{eq:BTZDil}
	\Phi = \Phi_r \frac{\sqrt{g_{\varphi\varphi}}}{\ell^2} = \frac{2\pi\Phi_r }{\beta} \frac{1}{\tanh \frac{\pi}{\beta}(u-v)}~.
	\end{equation}
Again, this matches the dilaton in \cite{AlmheiriOutside} up to the extremal contribution $\Phi_0$. 

\subsection{Boundary action}
In JT gravity, the boundary term in the action famously leads to the Schwarzian action. We wish to reproduce the Schwarzian action from the three-dimensional point of view. The Gibbons-Hawking term is 
	\begin{equation}
	\begin{aligned}
		S_{\rm GH} = \frac{1}{8 \pi G^{(3)}} \int d^2 x \sqrt{-h} \, \lp K^{(3)} + \frac{2}{\ell} \rp = \frac{2\pi \alpha \ell}{8\pi G^{(3)}} \int \dd t \sqrt{-h_{tt}} \, \phi_b  \lp K^{(3)} + \frac{2}{\ell} \rp~,
	\end{aligned}
	\end{equation}
where $\phi_b$ is the boundary value of $\phi$. The trace of the extrinsic curvature splits into two parts: 
	\begin{equation} \label{eq:dynK}
		K^{(3)} = h^{\mu\nu} K_{\mu\nu} = K^{(2)} + h^{\varphi\varphi} K_{\varphi\varphi}~. 
	\end{equation}
We evaluate the two-dimensional $K^{(2)}$ below and first focus on the contribution $h^{\varphi\varphi}K_{\varphi\varphi}$. To perform the dimensional reduction, we initially choose the boundary to be at a fixed value of $z$. We find both in Poincar\'e as well as in BTZ
	\begin{equation}
		h^{\varphi\varphi}K_{\varphi\varphi}  = - \frac{1}{\ell}~,
	\end{equation}
which is expected for the curvature of a circle. Thus, the boundary term in the action is (using $\Phi = 2\pi\alpha \phi$)
	\begin{equation} \label{eq:GHred}
		S_{\rm GH} = \frac{1}{8\pi G}\int \dif t \sqrt{-h_{tt}}\, \Phi_b \lp K^{(2)} + \frac{1}{\ell} \rp  ~. 
	\end{equation}
We will use this action to describe the dynamics. We will denote the boundary time coordinate with $t$, which becomes a parameter for the dynamical boundary trajectory $\lp \tau (t), z (t)\rp$  \cite{AlmheiriPolchinski, HermanBackreaction,SarosiSYK}. Here, $\tau$ and $z$ are (fixed) coordinates on the AdS$_2$ boundary; we will choose them to be the Poincar\'e coordinates. We then require that the boundary of AdS$_2$, i.e.\ the surface $u = v \equiv t$, coincides with the general boundary $X^+(u) = X^-(v)$. This defines the dynamical boundary time to be 
	\begin{equation}
		X^+(t) = X^-(t) \equiv \tau (t)~. 
	\end{equation}
We demand that the induced metric satisfies $g|_{bdy} = h_{tt} = - \frac{\ell^2}{\epsilon^2}$, which then implies that $z = \epsilon \sqrt{(\tau ')^2 - (z')^2} = \epsilon \tau ' + O(\epsilon^3)$. The normal to the boundary $z = z(t(\tau ))$ is 
	\begin{equation}
		n_a = \frac{\ell}{z} \frac{1}{\sqrt{\tau'^2 - z'^2}}(-z', \tau')~.
	\end{equation}
This gives $K^{(2)} \approx - \frac{1}{\ell} + \frac{\epsilon^2}{\ell} \{\tau, t\}$, with $\{\tau,t\} = \frac{\tau'''}{\tau'}- \frac{3}{2} \big( \frac{\tau''}{\tau'}\big)^2$ the Schwarzian derivative. Thus, the Gibbons-Hawking term evaluates to 
	\begin{equation} \label{eq:GHbdy}
		S_{\rm GH} =  \frac{1}{8\pi G} \int \frac{\dif t\, \ell}{\epsilon}\, \Phi_b \lp K^{(2)} +\frac{1}{\ell}  \rp =   \frac{1}{8\pi G}\int \dif t \, \Phi_r \{\tau (t), t\}~, 
	\end{equation}
where we defined $\Phi_b = \frac{\Phi_r}{\epsilon}$. As before, $\Phi$ is dimensionless such that $\Phi_r$ has dimensions of length. 

\section{Generalized entropy of JT black holes} \label{sec:entropies}
In this section we compute the generalized entropy for intervals in the extremal and finite temperature black hole plus bath systems. We do so from the higher-dimensional point of view discussed in the previous section, i.e.\ we will use geodesics.

\subsection{Extremal \texorpdfstring{AdS$_2$}{AdS2} black hole} \label{sec:extentropy}
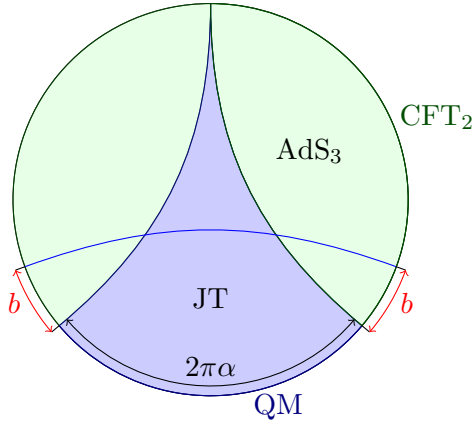
\begin{figure}
\begin{center}
\begin{tikzpicture}[scale=1.3]
	\draw (0,0) circle (2);
	\path (0,2) arc (90:220:2) coordinate (a);
	\filldraw[fill=green, draw=green!30!black, fill opacity=0.1, draw opacity = 0.8] (a) to[out=40, in=-90] (0,2) arc (90:220:2);
	\path (0,2) arc (90:320:2) coordinate (b);
	\filldraw[fill=green, draw=green!30!black, fill opacity=0.1, draw opacity = 0.8] (0,2) to[out=270, in=140] (b) arc (-40:90:2);
	\filldraw[fill=blue!40!white, draw=blue!50!black,  fill opacity = 0.5, draw opacity = 0.9] (a) to[out=40,in=-90] (0,2) to[out=270,in=140] (b) arc (-40:-140:2);
	\draw[blue!50!black] (b) arc (-40:-140:2) node[pos=.3, below] {QM};
	\draw[green!30!black] (0,2) to[out=270, in=140] (b) arc (-40:90:2) node[pos=.5, right] {CFT$_2$};
	\node at (0,-1) {JT};
	\node at (1, 0.5) {AdS$_3$};
	\path (0,0) --++ (-40:1.9) coordinate (c);
	\draw[<->] (c) arc (-40:-140:1.9) node[pos=.5, above] {$2\pi \alpha$};
	\path (a) arc (-140:-160:2) coordinate (geod1); 
	\path (b) arc (-40:-20:2) coordinate (geod2);
	\draw[blue] (geod1) to[out=20,in=160] (geod2);
	\draw (geod1) --++(-160:0.1) coordinate (geod1plus);
	\draw (geod2) --++(-20:0.1) coordinate (geod2plus);
	\draw (a) --++(220:0.1) coordinate (aplus);
	\draw (b) --++(320:0.1) coordinate (bplus);
	\draw[red, <->] (geod1plus) arc(-160:-140:2.1) node[pos=.5, left] {$b$};
	\draw[red, <->] (geod2plus) arc(-20:-40:2.1) node[pos=.5, right] {$b$};
\end{tikzpicture}
\caption{AdS$_3$-Poincar\'e, partially reduced over the angle $2\pi \alpha$. The purple region is the 2D JT extremal black hole, and the green region is dual to the bath 2D CFT. The blue geodesic computes the entropy of the region $[0,b]$ in the CFT, including the quantum mechanical system. \label{fig:geod1}}
\end{center}
\end{figure}

In \cite{AlmheiriOutside}, the generalized entropy for an interval in the extremal black hole + bath system was computed from the two-dimensional point of view. Here, instead, we want to compute the generalized entropy from the point of view of AdS$_3$. The setup that we have in mind is depicted in figure \ref{fig:geod1}.
Here, we have done a partial dimensional reduction of the $\varphi$-direction, i.e.\ instead of integrating the coordinate $\varphi$ in \eqref{eq:Poincarecylin} over $2\pi$ we integrated over some angle $2\pi \alpha$ with $\alpha \in (0,1]$. The result is that the spacetime has been split into two parts: one is the JT black hole, the other is dual to the CFT/bath system. Now, if we consider an interval $[0,b]$ in the CFT/bath system --- which also includes the quantum mechanical degrees of freedom --- its entropy will be given by the length of the blue geodesic in figure \ref{fig:geod1}. This can be easily computed using embedding coordinates. The details are in appendix \ref{sec:appext}. The entropy of an interval of which the endpoints lie at the same radial distance $r$ and separated by an angular interval $\Delta \varphi$ is given by 
	\begin{equation} 
		S = \frac{1}{2G} \asinh \frac{r \Delta \varphi}{2\ell}~. 
	\end{equation}
Here and in what follows $G = G^{(2)}$. If we take the endpoints to lie on the boundary, as is the case for the geodesic in figure \ref{fig:geod1}, one can expand to get
	\begin{equation}\label{eq:extentropy}
		S = \frac{1}{4G} \lp  2 \log \frac{\Phi_r + 2b}{\ell} \rp~, 
	\end{equation}
where we used $\Phi_r = 2\pi\ell \alpha$ and dropped the (constant) UV cutoff, as we will be interested in comparing different entropies.
 
\subsection{Finite temperature \texorpdfstring{AdS$_2$}{AdS2} black hole} \label{sec:finentropy}
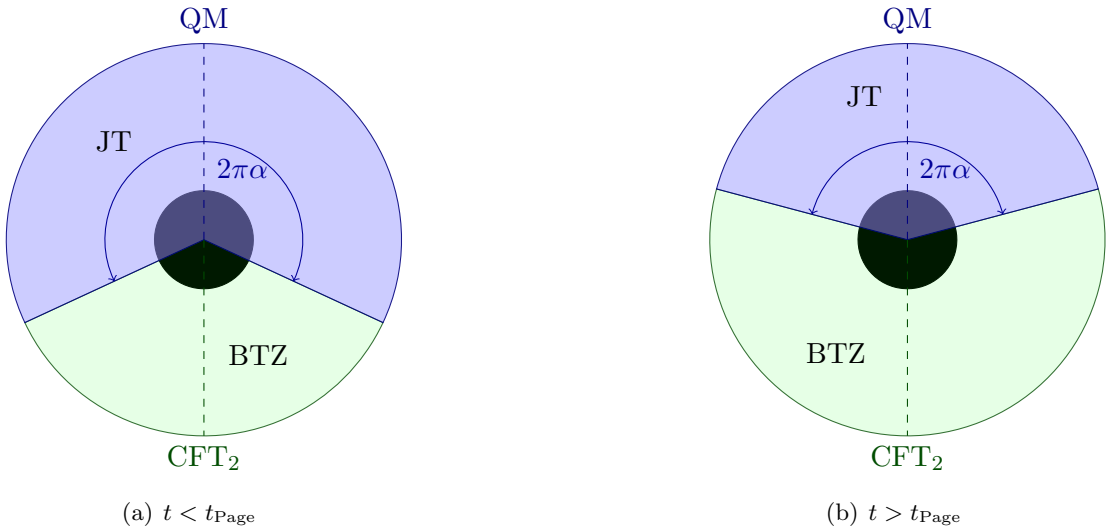
\begin{figure}
\centering
\subfigure[{$t < t_\text{Page}$} \label{fig:BTZJT:before}]
{
\begin{tikzpicture}[scale=1.3,rotate=115]
	\filldraw (0,0) circle (0.5cm);
	\filldraw[fill=green, draw=green!30!black, fill opacity = 0.1, draw opacity = 0.8] (0,0) -- (0,2) arc (90:220:2) -- cycle;
	\filldraw[fill=blue!40!white, draw=blue!50!black,  fill opacity = 0.5, draw opacity = 0.9] (0,0) -- (0,2) arc (90:-140:2) -- cycle;
	\draw[<->, blue!60!black] (0,1) arc (90:-140:1) node[pos=.6,below] {$2\pi \alpha$};
	\draw[blue!50!black, dashed] (0,0)--++(-25:2) node[pos=1,above] {QM};
	\draw[green!30!black, dashed] (0,0)--++(155:2) node[pos=1,below] {CFT$_2$};
	\node (JT2) at (1.3,0.4) {JT};
	\node (CFT2) at (-1.3,0){BTZ};
\end{tikzpicture}
}
\hfill
\subfigure[{$t>t_\text{Page}$} \label{fig:BTZJT:after}]
{
\begin{tikzpicture}[scale=1.3,rotate=75]	
	\filldraw (6,0) circle (0.5cm);
	\filldraw[fill=green, draw=green!30!black, fill opacity = 0.1, draw opacity = 0.8] (6,0) -- (6,2) arc (90:300:2) -- cycle;
	\filldraw[fill=blue!40!white, draw=blue!50!black,  fill opacity = 0.5, draw opacity = 0.9] (6,0) -- (6,2) arc (90:-60:2) -- cycle;
	\draw[<->, blue!60!black] (6,1) arc (90:-60:1) node[pos=.65,below] {$2\pi \alpha$};
	\draw[blue!50!black, dashed] (6,0) --++(15:2) node[pos=1, above] {QM};
	\draw[green!30!black, dashed] (6,0) --++(195:2) node[pos=1, below] {CFT$_2$};
	\node (JT4) at (7.3,0.8) {JT};
	\node (CFT4) at (4.7,0.4) {BTZ};
\end{tikzpicture}
}
\caption{We have done a partial reduction over the angle $2\pi\alpha$. The value of $\alpha$ determines if the black hole is before (a) or after (b) the Page time. \label{fig:BTZJT}}
\end{figure}

\begin{figure}
\centering
\subfigure[{$t<t_\text{Page}$} \label{fig:geodBTZ:1}]
{
\begin{tikzpicture}[scale=1.35, rotate=115]
	\filldraw (0,0) circle (0.5cm);
	\filldraw[fill=green, draw=green!30!black, fill opacity = 0.1, draw opacity = 0.8] (0,0) -- (0,2) arc (90:220:2) coordinate (start)  -- cycle;
	\filldraw[fill=blue!40!white, draw=blue!50!black,  fill opacity = 0.5, draw opacity = 0.9] (0,0) -- (0,2) arc (90:-140:2) -- cycle;
	\draw[blue!50!black, dashed] (0,0)--++(-25:2) ;
	\draw[green!30!black, dashed] (0,0)--++(155:2);
	\draw (0,2)--(0,2.1);
	\draw[<->, red] (0,2.05) arc (90:105:2.05) coordinate (end2) node[pos=.7, left] {$b$};
	\path (start) --++ (220:0.05) coordinate (start2);
	\draw[<->, red] (start2) arc (220:205:2.05) coordinate  (end) node[pos=.7, right] {$b$};
	\draw (start2) --++(40:0.05);
	\draw (end) --++(10:0.05) coordinate (blue1);
	\draw (end2) --++(300:0.05) coordinate (blue2);
	\draw (end2) --++(105:0.05);
	\draw (end) --++(205:0.05);
	\draw (start2) --++(220:0.05);
	\draw[blue] (blue1) to[out=25, in=285] (blue2);
\end{tikzpicture}
}
\hfill
\subfigure[{$t>t_\text{Page}$} \label{fig:geodBTZ:2}]
{
\begin{tikzpicture}[scale=1.3, rotate=75]
	\filldraw (0,0) circle (0.5cm);
	\filldraw[fill=green, draw=green!30!black, fill opacity = 0.1, draw opacity = 0.8] (0,0) -- (0,2) arc (90:300:2) -- cycle;
	\filldraw[fill=blue!40!white, draw=blue!50!black,  fill opacity = 0.5, draw opacity = 0.9] (0,0) -- (0,2) arc (90:-60:2) coordinate (Right) -- cycle;
	\path[name path={CFTline}] (0,2) arc (90:220:2);
	\path[name path={lineb}] (0,0) --++ (105:2.3);
	\draw[name intersections ={of=CFTline and lineb, by={Int1}}] (Int1) --++ (105:0.1cm);
	\draw (0,2)--(0,2.1);
	\draw[<->, red] (0,2.05) arc (90:105:2.05) node[pos=.5, left] {$b$};
	\path[name path={lineb2}] (0,0)--++(285:2.3);
	\path[name path={CFTline2}] (Right) arc (300:250:2);
	\draw[name intersections={of=CFTline2 and lineb2, by={Int2}}] (Int2) --++ (285:0.1cm);
	\draw (Right) --++(300:0.1cm);
	\path (Right) --++(300:0.05) coordinate (Red2);
	\draw[<->, red] (Red2) arc (300:285:2.05) node[pos=.5, right] {$b$};
	\path[blue] (0,0.7) arc (90:15:0.7) coordinate (halfway);
	\draw[blue] (Int1) to[out=285,in=105] (halfway);
	\draw[blue] (halfway) to[out=285,in=105] (Int2);
	\draw[dashed] (0,0) --++(15:2);
	\draw[dashed] (0,0) --++(195:2);
\end{tikzpicture}
}
\caption{To find the entropy of the double interval $[0,b]$, including the quantum mechanical degrees of freedom, we need to compute the length of the blue geodesics. \label{fig:geodBTZ}}
\end{figure}
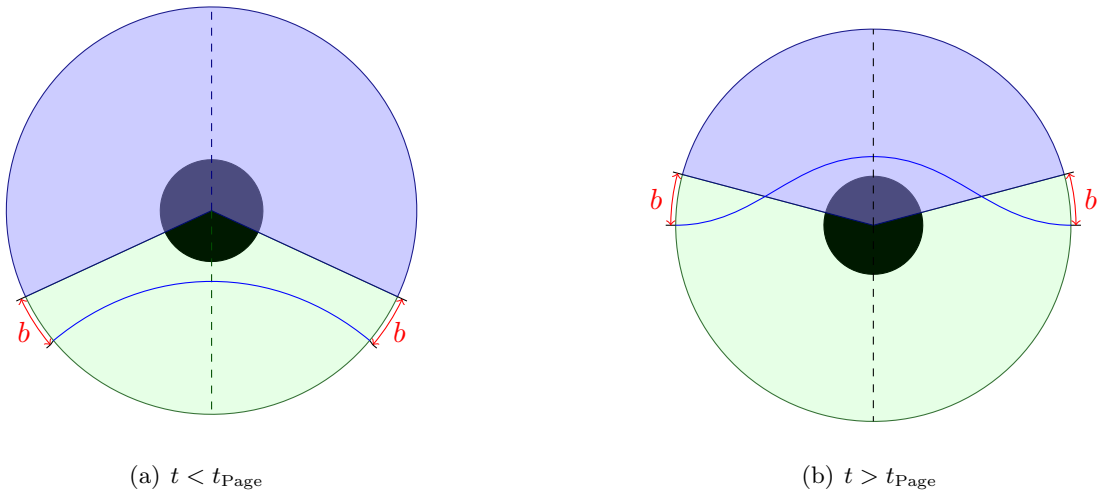

For the BTZ case, the setup is as in figure \ref{fig:BTZJT}. 
Again, we have done a partial dimensional reduction over the $\varphi$-coordinate up to $2\pi \alpha \in (0, 2\pi]$. The corresponding region in the BTZ black hole now reduces to a black hole in JT gravity (purple region). The remainder (green region) we view as dual to the bath (2D CFT) degrees of freedom. By decreasing the value of $\alpha$ from $1$ to $0$, we can geometrically `evaporate' the black hole. We will comment more on this in the next section, in which we discuss the dynamics of our model. For now, we will distinguish two cases: `before' and `after' the Page time or the half-way evaporation point (figure \ref{fig:BTZJT:before} and \ref{fig:BTZJT:after}, respectively). In both cases, the entropy of an interval $[0, b]$ in the CFT, which also includes the quantum mechanical degrees of freedom, is given by the length of a geodesic in BTZ of which the endpoints lie at the boundary (on a fixed time slice). The details are in appendix \ref{sec:appfinite}. Before taking the endpoints to the boundary, the entropy of such an interval is given by 
	\begin{equation}
		S = \frac{1}{2G} \asinh \frac{2\pi \ell^2 r}{\beta} \sinh{\frac{\pi}{\beta} \ell \Delta \varphi}~, 
	\end{equation}
where $\Delta \varphi$ is the angular separation of the two points. For the geodesic in figure \ref{fig:geodBTZ:1}, i.e.\ before the Page time, we then find (expanding for $r \to \infty$)
	\begin{equation} \label{eq:entbeforePage}
		S = \frac{1}{4G}\Big(2 \log{\sinh{\frac{\pi}{\beta}\lp 2\pi \ell (1- \alpha ) -  2b \rp}} \Big)~, 
	\end{equation}
where again we dropped the UV cutoff. After the Page time, the geodesic `jumps' and crosses the purple region (see figure \ref{fig:geodBTZ:2}). Thus the entropy is now given by 
	\begin{equation} \label{eq:entafterPage}
		S = \frac{1}{4G} \Big( 2\log{\sinh{ \frac{\pi}{\beta} \left(2\pi \ell \alpha + 2b\right) }}\Big)~.
	\end{equation}
Eventually, we are interested in the entropy of the entire bath of radiation, i.e.\ we wish to take $b \to 0$. We will do so in section \ref{sec:entropyPage}.

Notice that we are viewing the full black hole + bath geometry as being in a pure state: we think of the BTZ black hole as the result of some collapsing matter, i.e.\ the resulting three-dimensional geometry is still describing a pure state. Since we are interested in obtaining a Page curve, this is also the natural state to consider. If we instead consider the geometry to be in the thermal state, we have to add a second, disconnected contribution (the BTZ black hole area) to the entropy if the angular interval grows larger than some critical angle \cite{OoguriBao}. In that case, the transition occurs much later than the halfway evaporation point and one would not reproduce the Page curve. In fact, the inclusion of the black hole entropy would lead to an entanglement entropy that is more reminiscient of the Hawking curve. 

\section{Dynamical evaporation} \label{sec:dynamics}
The black holes we have discussed so far will not dynamically evaporate, because the black hole is in thermal equilibrium with the CFT. 
This means that the in- and out-flux of energy are equal.  If we allow instead for particles to escape, we can simulate an evaporation process. Our model is somewhat similar to the recent models \cite{Penington1,AlmheiriEMM,AlmheiriMMZ} (see also e.g.\ \cite{Hollowood1, Hollowood2}, in which Hawking radiation is allowed to escape to the `bath' consisting of a 2D CFT on the half-line). Note that in some other models, such as \cite{AlmheiriOutside, MyersEquil}, the black hole is in equilibrium with the bath. In that case, the black hole does not evaporate and an island appears outside of the horizon. The model we propose lies somewhere in between. For the purposes of discussing the evaporation process, we will only consider the finite temperature black hole. 

In our model, the JT gravity theory is obtained by a dimensional reduction of the three-dimensional AdS gravity theory on the BTZ background. One can also view the JT gravity theory as being the induced theory on a tensionless end-of-the-world (EOW) brane that bounds the BTZ geometry. Note that, unlike in other brane world scenarios, there is no explicit gravitational action added on the boundary. In this sense, the 2D JT gravity can be seen as being equivalent to the bulk theory that has been removed from the BTZ geometry. Indeed, in the microscopic description we identified the AdS$_2$ black hole as being represented by the removed part of the full quantum state of the boundary theory (i.e.\ the purple part of the boundary in figures \ref{fig:BTZJT} and \ref{fig:geodBTZ}). Note that this part is not sufficient to describe the entire integrated-out bulk region, but only describes the entanglement wedge bounded by the geodesic in figure \ref{fig:geodBTZ}. The remaining part is described by the bath. 

The induced JT gravity theory couples at the boundary to the CFT bath. To model the evaporation process we allow a  small amount of energy to leak in to the CFT bath. From the microscopic perspective there is no mystery in this evaporation process: by adjusting the temperature of the bath to be slightly below that of the QM system, there will always be a net heat flow into the bath. In the gravity dual this energy flux is interpreted as Hawking radiation that has escaped the black hole. Note that for this process to occur it is not essential that the dual theory contains local matter fields. Even the pure gravitational degrees of freedom are capable of transferring energy from the black hole to the boundary. 

In the preceding sections, we stated that we can consider different phases of evaporation by means of the parameter $\alpha$, which controls the dimensional reduction (and is closely related to the dilaton). So far, we did not discuss a dynamical way of changing ${\Phi_r \sim \alpha}$. In this section, we will add explicit time-dependence to the renormalized dilaton ${\Phi_r (t) = 2\pi \ell \alpha (t)}$ and see that this indeed results in an evaporating black hole. Then, we will map this to a more standard evaporation protocol in which the dilaton is fixed, but the temperature (and mass) of the black hole decrease. To do so, we will use a mapping between two BTZ geometries with different parameters. We will first briefly discuss this mapping, and then proceed to compare the two perspectives on evaporation. 

\subsection{Mapping two BTZ geometries} \label{sec:BTZmap}
We consider a map between two BTZ geometries with different parameters introduced in \cite{deBoerBTZ}: 
	\begin{equation}
		\text{BTZ}(M\lambda^2;2\pi) \equiv \text{BTZ}(M;2\pi\lambda)~.
	\end{equation}
On the left hand side we have a BTZ geometry with a mass that can vary; on the right hand side we have a BTZ geometry with a varying conical deficit. We start from the usual BTZ metric
	\begin{equation} \label{eq:BTZmetric}
		ds^2 = - \lp \frac{r^2}{\ell^2} - \frac{4\pi^2 \ell^2}{\beta^2}\rp \dif t^2 + \lp \frac{r^2}{\ell^2} - \frac{4\pi^2 \ell^2}{\beta^2}\rp^{-1} \dif r^2 + r^2 \dif \varphi^2~,
	\end{equation}
where $\ell$ is the AdS length, the horizon is at $R^2 = 8GM\ell^2$ and $\varphi \sim \varphi + 2\pi$ is identified. The inverse temperature is $\beta = \frac{2\pi \ell^2}{R}$. Now, consider the transformation
	\begin{equation}
	\begin{aligned}
		r = \lambda \tilde{r}, \quad R = \lambda \tilde{R}, \quad t = \lambda^{-1} \tilde{t}, \quad \varphi = \lambda^{-1} \tilde{\varphi}~. 
	\end{aligned}
	\end{equation}
This keeps the form of the metric invariant as in \eqref{eq:BTZmetric}, but the periodicity in $\tilde{\varphi}$ is now $2\pi \lambda$.  Under this $\lambda$-transformation, the entropy $S = \frac{2\pi R}{4G^{(3)}}$ remains invariant, but the Hawking temperature $T_H = \frac{R}{2\pi \ell^2}$ gets scaled by $\lambda^{-1}$.  If we pick $\lambda = \frac{2\pi \ell}{\beta}$ and leave $R$ (or equivalently $\beta$) fixed, we find 
	\begin{equation} \label{eq:BTZtilded}
		ds^2 = - \lp \frac{\tilde{r}^2}{\ell^2}-1\rp \dif\tilde{t}^2 + \lp \frac{\tilde{r}^2}{\ell^2}-1 \rp^{-1} \dif \tilde{r}^2 + \tilde{r}^2 \dif\tilde{\varphi}^2~.
	\end{equation}
In what follows, we will use tilded coordinates to describe the BTZ geometry on which we performed a partial reduction in section \ref{sec:finentropy}. Indeed, in this case the temperature is fixed and the periodicity of $\tilde{\varphi}$ leads us to identify $\alpha$ with $\lambda = \frac{2\pi \ell}{\beta}$. 
We will study the dynamics of this model in the next section. We will then use the above map in section \ref{sec:lintoexp} to find untilded coordinates in which the black hole energy decays exponentially. 

\subsection{Black hole evaporation using boundary dynamics}\label{sec:BHevap}
To study the dynamics of our model, consider again the Schwarzian action 
	\begin{equation}
		S = \frac{1}{8\pi G} \int \dif t\,\Phi_r \{ \tau, t\}~, 
	\end{equation}
where we will allow for $\Phi_r$ to be time-dependent. Varying with respect to $\tau (t)$ gives the equation of motion (in the absence of matter terms)
	\begin{equation} \label{eq:EOM}
		 \frac{1}{\tau '} \lp \Phi_r \{\tau, t\}' + 2\Phi_r' \{\tau, t\} + \Phi_r''' \rp = 0~, 
	\end{equation}
where primes denote $t$-derivatives. If $\Phi_r$ is constant, this reduces to the more familiar 
	\begin{equation}
		\frac{1}{\tau'}\{\tau, t\}' = 0~, 
	\end{equation}
i.e.\ we are looking for non-constant functions $\tau (t)$ with constant Schwarzian. This leads to the solution for the non-evaporating black hole, where 
	\begin{equation}
		\tau = \frac{\beta}{\pi} \tanh{\frac{\pi}{\beta} t}~, 
	\end{equation}
with constant inverse temperature $\beta$ and (constant) ADM energy 
	\begin{equation} \label{eq:ADM}
		E = - \frac{\Phi_r}{8\pi G} \{\tau, t\} = \frac{2\pi^2}{\beta^2} \frac{\Phi_r^0}{8\pi G} \equiv E_0~. 
	\end{equation}
Now we will restore time-dependence and use tilded quantities to distinguish from the above case. Remember that in the model at hand, we have done a partial reduction of the BTZ geometry, resulting in a JT gravity part with dilaton 
	\begin{equation}
		\tilde{\Phi}_r = 2\pi \ell \alpha (\tilt) \equiv \Phi_r^0 \alpha (\tilt)~, 
	\end{equation}
where we defined $\Phi_r^0 \equiv 2\pi \ell$ and $\alpha (\tilt)$ decreases from 1 to 0. We interpret the remaining part as holographically dual to a 2D CFT. Now, the decreasing $\alpha$ does not affect the temperature of the 2D black hole, which is directly inherited from the BTZ black hole. Therefore, we will consider this temperature to be fixed and we will keep $\tau = \frac{\beta}{\pi} \tanh{\frac{\pi}{\beta} \tilt}$. To represent the interaction with the bath, we add an extra term to the equation of motion \eqref{eq:EOM}, equal to the incoming minus the outgoing energy flux (c.f. \cite{HermanBackreaction, AlmheiriEMM}).  
	\begin{equation} \label{eq:EOM2}
		 - \frac{1}{8\pi G}\lp \tilde{\Phi}_r \{\tau, \tilt \}' + 2\tilde{\Phi}_r' \{\tau, \tilt\} + \tilde{\Phi}_r''' \rp = \tilde{T}_{vv}(\tilt) - \tilde{T}_{uu}(\tilt) =~ :\tilde{T}_{vv}(\tilt): - :\tilde{T}_{uu}(\tilt):~. 
	\end{equation}
Initially, we are in thermal equilibrium and the right hand side vanishes. At $\tilt = 0$ we break thermal equilibrium, such that there is a net outgoing energy-momentum flux on the boundary for $t > 0$, 
	\begin{equation} \label{eq:netflux}
		:\tilde{T}_{uu} (\tilt):\, = - \frac{c}{24\pi} \{\tau, \tilt\}~, \qquad :\tilde{T}_{vv} (\tilt):\, = -(1-\epsilon ) \frac{c}{24\pi} \{\tau, \tilt\}~.
	\end{equation}
Note that perfect absorbing boundary conditions mean $\epsilon = 1$, such that $:\tilde{T}_{vv} (\tilt ): = 0 $; here, we are interested in $\epsilon \ll 1$, such that the evaporation is adiabatic.
One can think of this net flux as simply the effect of moving the dividing line between the JT gravity and CFT part of our BTZ black hole, as in figure \ref{fig:BTZJT}: we are relabeling which degrees of freedom are `in' and which are `out' of the black hole. Then, since $\{\tau, \tilt\} = - \frac{2\pi^2}{\beta^2} = \text{cst}$, \eqref{eq:EOM2} gives
	\begin{equation}
		- \frac{1 }{4\pi G} \Phi_r'  =  \epsilon \frac{c}{24\pi}~, 
	\end{equation}	 
which we can solve to find 
	\begin{equation}
		\tilde{\Phi}_r = \Phi_r^0\, \alpha (\tilt ) = \Phi_r^0 \lp 1 - \frac{A}{2} \tilt \rp, \quad \text{where} \quad  \frac{A}{2} = \epsilon  \frac{c}{6} \frac{G}{\Phi_r^0} ~.
	\end{equation}
We think of $\frac{A}{2}$ as an evaporation rate (we pick the factor $\frac{1}{2}$ for later convenience). In terms of this evaporation rate, the energy in this coordinate system decreases as
	\begin{equation}\label{eq:solA}
		\frac{d\tilde{E}}{d \tilt} = \tilde{T}_{vv} -\tilde{ T}_{uu} = -\epsilon \frac{c}{24\pi} \frac{2\pi^2}{\beta^2} = - E_0 \frac{A}{2}~.
	\end{equation}
Here and below, $\beta$ should be interpreted as a constant, unless explicitly written as $\beta (t)$. 

\subsection{From linear to exponential evaporation} \label{sec:lintoexp}
In many models of black hole evaporation, the energy decreases exponentially in time (see e.g.\ \cite{HermanBackreaction, AlmheiriEMM, Hollowood1, Hollowood2}). In those models, the two-dimensional JT black hole is put into contact with an external bath. We will now show how our model relates to this type of evaporation, in which the temperature depends on time and the dilaton is fixed. To do so, we will exploit the mapping discussed in section \ref{sec:BTZmap}.

As a first step, we need to find $\tilt (t)$. A simple solution comes from considering again the Schwarzian action with constant dilaton and changing $t \to \tilt (t)$: 
	\begin{equation}
	\begin{aligned}
		S &= \frac{1}{8\pi G} \int \dd t\, \Phi_r^0 \{\tau, t\} = \frac{1}{8\pi G} \int \dd t \, \Phi_r^0 \Big[ \Big(\frac{dt}{d\tilt} \Big)^{-2} \{\tau, \tilt\} - \{\tilt, t\} \Big] \\
		&= \frac{1}{8\pi G} \int \dd \tilt \, \Phi_r^0 \Big(\frac{d\tilt}{d t} \Big) \{\tau, \tilt\} = \frac{1}{8\pi G} \int \dd \tilt \, \tilde{\Phi}_r (t) \{\tau, \tilt\}
	\end{aligned}
	\end{equation}
where in the second line we assumed that $\{\tilt, t\}$ is constant, as we will later confirm. Then requiring that the action is invariant leads us to conclude that $\tilde{\Phi}_r (t) = \Phi_r^0 \frac{d\tilt}{d t} $, i.e. 
	\begin{equation} \label{eq:alphatilde}
		\frac{d \tilt}{d t} = \alpha (\tilt) = 1 - \frac{A}{2} \tilt~. 
	\end{equation}
Solving \eqref{eq:alphatilde}, we find that $t$ and $\tilt$ are related by 
	\begin{equation} \label{eq:ttilde}
		\tilde{t} (t) = \frac{2}{A} (1- e^{- A t/2}) \quad \Rightarrow \quad \frac{d \tilt}{d t}  = e^{-A t/2}~, \quad \{\tilt, t\} = - \frac{1}{2} \lp \frac{A}{2} \rp^2~.
	\end{equation}
For a similar discussion on exponential evaporation with a more complicated solution for $\tilde{t} (t)$, see \cite{Hollowood1,Hollowood2,MyersEquil}. We expect that our solution is a good approximation to these more general models in the limit of slow evaporation. 

\subsection{Exponentially evaporating black hole} \label{sec:expevap}
We are now ready to connect the linear/geometric evaporation presented in section \ref{sec:BHevap} to the more common exponential decay. As before, we start from the energy flux equation, which for constant dilaton reduces to
	\begin{equation}
		- \frac{1}{8\pi G} \Phi_r^0 \{ \tau, t\}' =\,: T_{vv}(t): - :T_{uu}(t):~.
	\end{equation}
A priori we do not know the energy flux on the right hand side. However, transforming \eqref{eq:EOM2} gives information on this. Since the right hand side is manifestly a tensor ($\tilde{T}_{\tilt\tilt}$) both the left and right hand side should transform as a tensor. Indeed we find that under $\tilt \to \tilt (t)$ we get
	\begin{equation}
		- \Big( \frac{d\tilt}{d t} \Big)^{-2} \lp \Phi_r \{\tau, t\}' + 2\Phi_r' \{\tau, t\} + \Phi_r''' \rp = \Big( \frac{d \tilt}{d t} \Big)^{-2} 8\pi G (T_{vv} - T_{uu})
	\end{equation}
Here, we have not yet assumed that $\Phi_r = \text{cst}$; we only assumed that it transforms as a vector, i.e.\ $\tilde{\Phi}_r (\tilt) = \frac{d \tilt}{d t} \Phi_r (t)$. Hence we conclude
	\begin{equation} \label{eq:transformedEOM}
	\begin{aligned}
		T_{vv} - T_{uu} = \lp \frac{d\tilt}{d t} \rp^2 (\tilde{T}_{vv} - \tilde{T}_{uu} ) = \lp \frac{d \tilt}{d t} \rp^2 \epsilon \frac{ c}{24\pi} \{\tau, \tilt\} ~.
	\end{aligned}
	\end{equation}
Thus, where we previously had a constant net energy flux, this time we have an exponentially decreasing net flux. Therefore, we conclude that the ADM energy decreases as
	\begin{equation}
		E = - \frac{\Phi_r}{8\pi G} \{\tau, t\} = E_0 e^{- A t} + \frac{1}{2}\Big(\frac{A}{2}\Big)^2 \frac{\Phi_r^0}{8\pi G}~,
	\end{equation}
and
	\begin{equation}
		\frac{d E}{d t} = - \frac{\Phi_r}{8\pi G}  \{\tau, t\}' = -\frac{c}{24\pi} \frac{2\pi^2}{\beta^2} e^{-A t}~.
	\end{equation}
Finally, from the map discussed in section \ref{sec:BTZmap} we know that
	\begin{equation}  \label{eq:exptemp}
		\frac{d \tilt}{d t}   = \frac{2\pi \ell}{\beta (t)} \quad \Rightarrow \quad \beta (t) = 2\pi \ell\, e^{\frac{A}{2} t}~.
	\end{equation}
As also noted in \cite{HermanBackreaction}, for low evaporation rates, i.e.\ in the regime $ A  \ll \frac{1}{\ell}$, the evaporation is adiabatic. The length scale $\frac{\Phi_r^0}{c \,\epsilon}$ sets the evaporation time of the black hole.

\subsection{Obtaining the Page curve}\label{sec:entropyPage}
From the Hawking temperature \eqref{eq:Htemp} we see that the entropy and energy are related via
	\begin{equation}
		S_\text{BH} = 2\pi \sqrt{\frac{E \Phi_r}{4\pi G}}~. 
	\end{equation}
Consider this formula in the tilded coordinate system discussed in section \ref{sec:BHevap}. We have
	\begin{equation} \label{eq:Cardyentropy}
		\tilde{S}_\text{BH} (\tilt ) = 2\pi \sqrt{\frac{\tilde{E}(\tilt) \tilde{\Phi}_r (\tilt ) }{4\pi G}} = \frac{2\pi}{\beta} \frac{\Phi_r^0}{4G} (1 - \frac{A}{2} \tilt )~, \quad \text{ where } \, \frac{A}{2} = \epsilon \frac{cG}{6\Phi_r^0} ~.
	\end{equation}
Hence this entropy decreases linearly in time $\tilt$ (remember that $\beta$ is fixed). Note that also via this method, we do not obtain the extremal entropy $S_0 = \frac{1}{4G} \Phi_0$, consistent with our earlier statement that the topological piece $\Phi_0$ is not relevant for our model. In the untilded coordinate system discussed in \ref{sec:expevap}, we find
	\begin{equation}
		S_\text{BH} (t) = 2\pi \sqrt{\frac{E(t)\Phi_r}{4\pi G}} = \frac{\Phi_r^0}{4G} \sqrt{\frac{A^2}{4} + \frac{4\pi^2}{\beta^2} e^{-At}}~.
	\end{equation}
\begin{figure}[t!]
\centering
	\begin{tikzpicture}[scale=1.1]
	\draw[->] (0,0) -- (6.5,0) node[below] {$\tilt$};
	\draw[->] (0,-0.1) -- (0,4.5) node[left] {$S$};
	\draw[orange] (0,0) -- (6,4);
	\draw[blue] (0,4) -- (6,0);
	\draw (0,4)--(-0.1,4) node[left] {\small{$\frac{1}{4G}\frac{2\pi \Phi_r^0}{\beta}$}}; 
	\draw (0,0)--(-0.1,0); 
	\draw (6,0)--(6,-0.1)  node[below] {\small{$2/A$}};
	\draw[red] (0.075,0) -- (3,1.95) -- (5.925,0);
	\draw[orange] (7,4) -- (7.5,4) node[right] {\color{black}$S_\text{rad}^{t<t_\text{Page}}$};
	\draw[blue] (7,3.3) -- (7.5,3.3) node[right] {\color{black}$S_\text{rad}^{t>t_\text{Page}}, S_{BH}$};
	\draw[red] (7,2.6) -- (7.5,2.6) node[right] {\color{black}$S_\text{rad}$};
	\end{tikzpicture}
	\caption{The entropy of the radiation follows a Page curve (red line).\label{fig:Pagecurve}}
\end{figure}
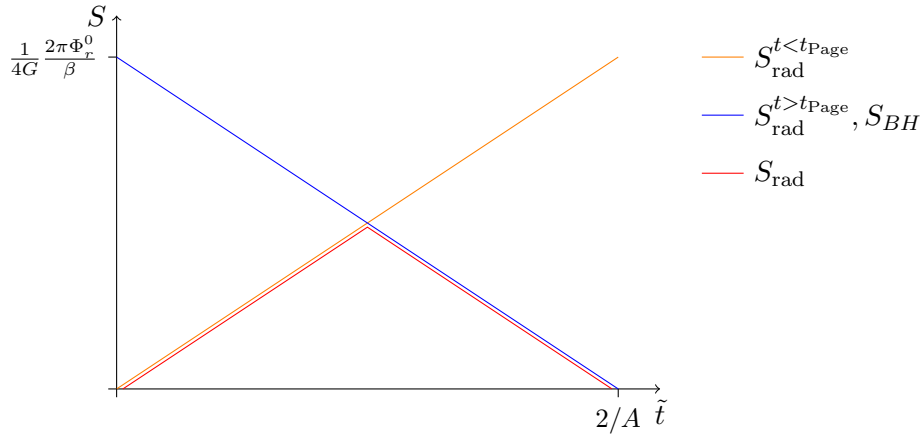
Next, we would like to make contact with our entropy calculations performed in section \ref{sec:entropies}. The comparison is most natural in the tilded coordinates. In section \ref{sec:finentropy} we found \eqref{eq:entbeforePage} and \eqref{eq:entafterPage} before and after the Page time, respectively. We are now interested in the entropy of the entire bath, i.e.\ we wish to take $b \to 0$. Then before the Page time we find
	\begin{equation}
		S^{\tilt < \tilt_\text{Page}} = \frac{1}{4G} \lp 2\log \sinh \frac{\pi}{\beta} \Phi_r^0 \frac{A}{2} \tilt \rp~. 
	\end{equation}
After the Page time we get
	\begin{equation}
		S^{\tilt > \tilt_\text{Page}} = \frac{1}{4G} \lp  2\log\sinh\frac{\pi}{\beta} \Phi_r^0 (1 - \frac{A}{2}\tilt) \rp~.
	\end{equation}
For high temperatures $\Phi_r^0 \gg \beta$ we can approximate this with
	\begin{equation} \label{eq:Pagecurve}
	S \approx \begin{cases} \frac{1}{4G} \lp  \Phi_r^0  \frac{2\pi}{\beta}  \frac{A}{2}\tilt \rp & \mbox{if } \tilt < \tilt_\text{Page} \\ \frac{1}{4G} \lp  \Phi_r^0  \frac{2\pi}{\beta}  (1 - \frac{A}{2} \tilt) \rp & \mbox{if } \tilt > \tilt_\text{Page} \end{cases} ~. 
	\end{equation}
Notice that for $\tilt > \tilt_\text{Page} = A^{-1}$ the entropy of the radiation is equal to the entropy of the black hole in \eqref{eq:Cardyentropy}. The latter is the coarse-grained entropy and follows a Hawking curve. From \eqref{eq:Pagecurve} it is clear that we indeed reproduce a Page curve for the radiation entropy; see figure \ref{fig:Pagecurve}. 

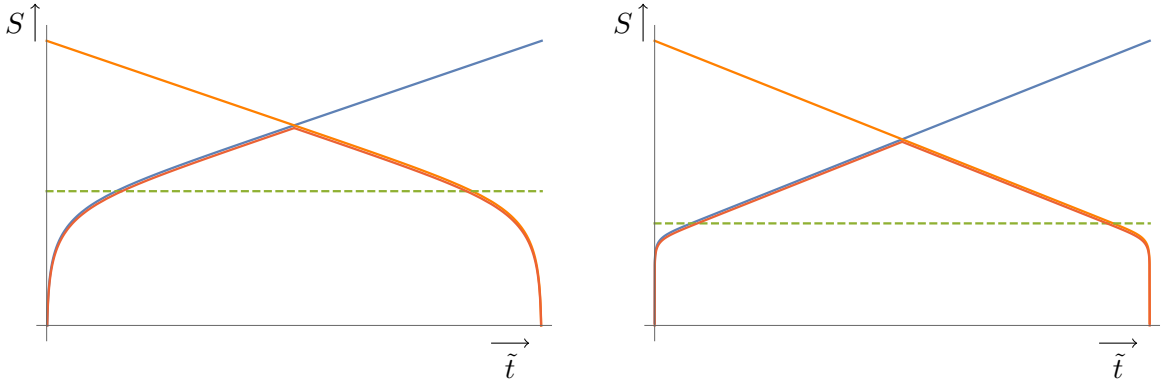
\begin{figure}[t!]
\centering
\begin{tikzpicture}
	\node[anchor=south west, inner sep=0] at (0,0) {\includegraphics[width=0.45\textwidth]{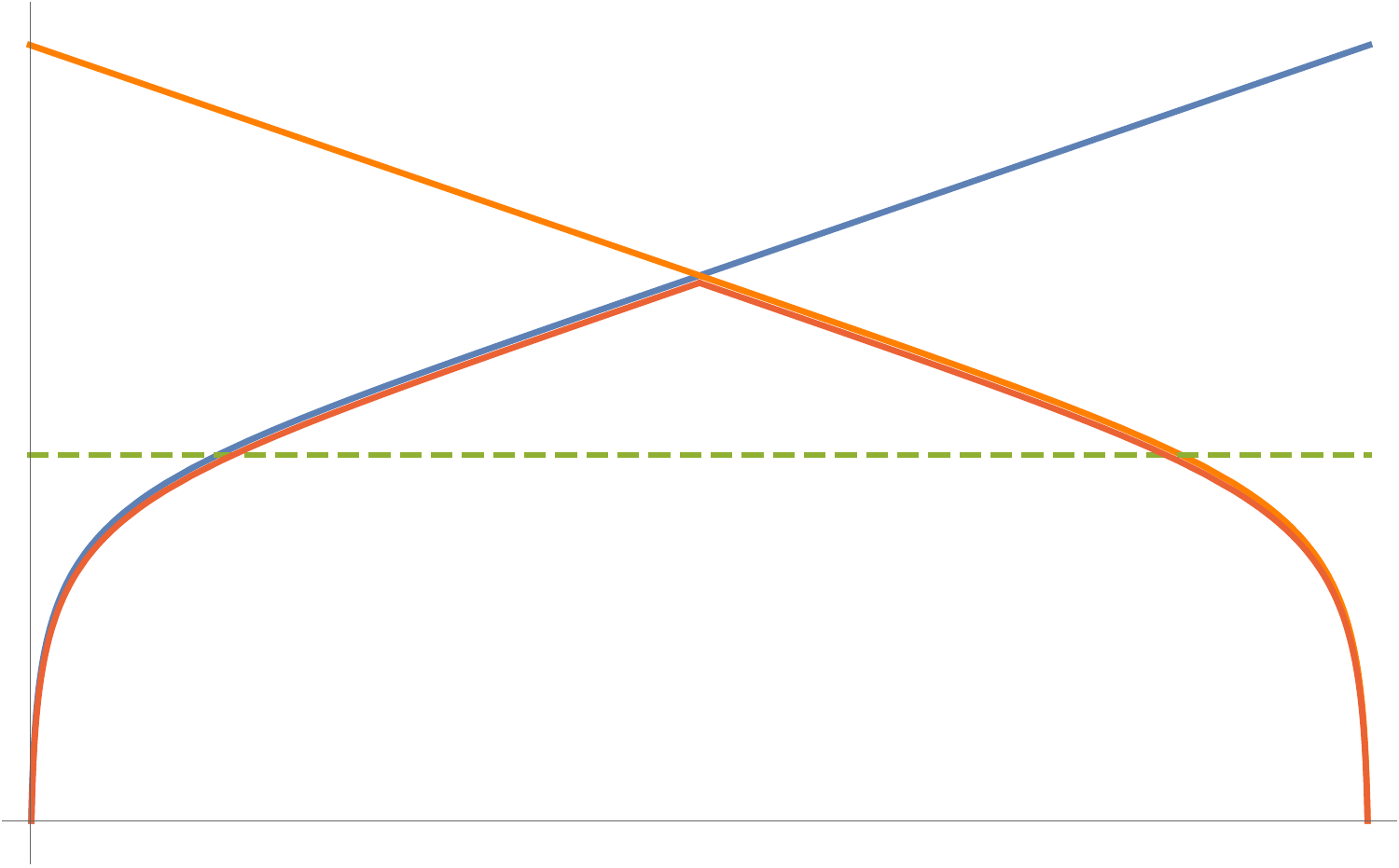}};
	\draw[->] (0,4)--(0,4.5) node[pos=.5,left] {$S$};
	\draw[->] (6,0)--(6.5,0) node[pos=.5,below] {$\tilt$};
	\node[anchor=south west, inner sep=0] at (8,0) {\includegraphics[width=0.45\textwidth]{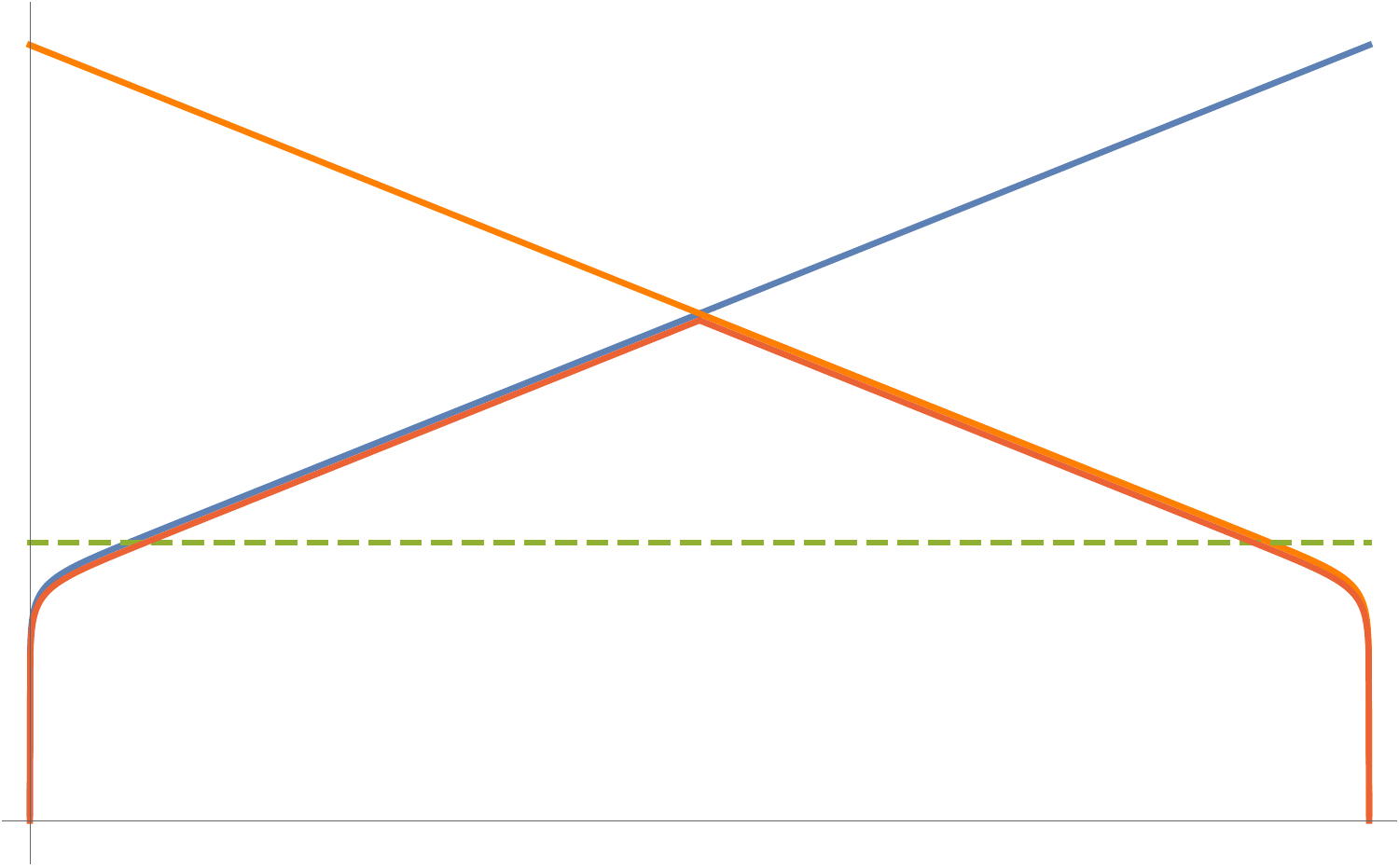}};
	\draw[->] (8,4)--(8,4.5) node[pos=.5,left] {$S$};
	\draw[->] (14,0)--(14.5,0) node[pos=.5,below] {$\tilt$};
\end{tikzpicture}
	\caption{The qualitative behavior of the entropy for $\Delta \varphi = 2\pi \frac{A}{2}\tilt$ (blue) and $\Delta \varphi = 2\pi ( 1 - \frac{A}{2} \tilt )$ (orange) shows that it follows the Page curve (red line). To plot we set $\ell = G = 1$, $r_{\infty} = 10^{100.000}$, the evaporation rate $A = 5$ and $R = 2 \ell$ on the left, $R = 10 \ell$ on the right. The dashed green line indicates the UV cutoff.\label{fig:arcsinh}}
\end{figure}

If we do not expand for large cutoff, and without zooming in on the regime $\Phi_r^0 \gg \beta$, we can study the qualitative behavior of the entropy using 
	\begin{equation}\label{eq:arcsinhentropy}
		S = \frac{1}{2G} \asinh{\frac{r_{\infty}}{R} \sinh{\frac{R}{2} \Delta \varphi}}~, 
	\end{equation}
where $R = \frac{2\pi \ell^2}{\beta}$ is the BTZ radius, $r_{\infty}$ is the cutoff surface, and $\Delta \varphi = 2\pi \frac{A}{2}\tilt$ before and $\Delta \varphi = 2\pi \lp 1 - \frac{A}{2} \tilt \rp$ after the Page time. This gives figure \ref{fig:arcsinh}. 

\section{Discussion}
In this paper, we investigated JT black holes of zero and finite temperature coupled to a bath (a 2D CFT) from a higher-dimensional, geometrical perspective. By performing a partial dimensional reduction from Poincar\'e AdS$_3$ and the BTZ geometry, respectively, we effectively split the three-dimensional spacetime into a two-dimensional black hole and a remainder, of which the holographic dual takes on the role of the bath. The boundary conditions on the dilaton lead us to identify the renormalized value of the dilaton with the parameter $\alpha$ controlling the dimensional reduction: $\Phi_r = 2\pi \ell \alpha$. This procedure allowed us to compute the entropy of an interval in the bath/radiation by simply computing geodesic lengths in the three-dimensional spacetime.

By making the dimensional reduction parameter $\alpha$ time-dependent, we could model the dynamics of the BTZ system and allowed the finite temperature JT black hole to evaporate. From a boundary analysis we showed that the energy decreases linearly in time. The renormalized dilaton is time-dependent and the temperature is fixed. Then, we exploited a mapping of BTZ parameters to connect this linear evaporation to a more canonical (exponential) evaporation of the energy, in which the renormalized dilaton takes on a constant value, and the temperature is time-dependent. Finally, we demonstrated that the entropy of the radiation/bath system follows a Page curve. 

We included the extremal black hole mostly as a toy model to study the dimensional reduction and demonstrate the entropy calculations. We did not include time dependence for the extremal black hole, because this black hole (which has $T_H = 0$) does not evaporate.  \\

\noindent A few comments are in place. First, the connection between the dilaton and the  black hole entropy naturally arises from our description in a geometrical fashion. This is best seen for the finite temperature case. From \eqref{eq:Cardyentropy} we see that the black hole entropy is given by the value of the dilaton \eqref{eq:BTZDil} at the horizon: 
	\begin{equation}
		S = \frac{1}{4G}\Phi \big|_\text{hor} = \frac{1}{4G}\lp \frac{2\pi}{\beta} \Phi_r \rp~.
	\end{equation}	 
Note that the entropy \eqref{eq:Cardyentropy} is a thermal and coarse-grained entropy; indeed, inserting $\Phi_r (\tilt ) = 2\pi \ell (1 - \frac{A}{2} \tilt )$ does not lead to the Page curve, but instead a linearly decreasing (Hawking) curve. Since we wish to consider the full geometry to be in a pure state, the fine-grained entropy of the JT black hole is equal to that of the radiation (its complement) and thus follows a Page curve as well. 

As a second comment on our results, note that we did not have to make use of the island formula to reproduce the Page curve for the radiation. Instead, we naturally find the Page curve from the RT prescription, which instructs us to take the minimal length geodesic in the BTZ geometry. As the interval on the boundary corresponding to the radiation grows, the geodesic `jumps' and its length starts to decrease. In the two-dimensional theory, this is mirrored by a jump of the quantum extremal surface, thereby including an island in the generalized entropy calculation. One might like to interpret the region in figure \ref{fig:geodBTZ:2} bounded by the geodesic and the division between the purple and green regions as the island. This island lies outside of the horizon, as expected for our adiabatic evaporation. It would be interesting to make this connection more precise. 

Finally, we would like to point out that the method we have described in this paper, i.e.\ using a partial dimensional reduction to create a black hole and bath within the same higher-dimensional system, could in principle be applied to other spacetimes as well. In particular, it could be worthwhile to apply this procedure to a three-dimensional Schwarzschild-de Sitter black hole, reducing it to a pure dS$_2$ spacetime connected to a bath, to see if we can learn more about the (pure) de Sitter horizon and a possible information paradox. So far, the literature has not been conclusive on the existence of islands in pure de Sitter (see e.g.\ \cite{Sybesma, CosmologyIslands, BalasubramanianDSislands, dSinfo, WatseLars}), leaving the matter of both the existence as well as the interpretation of a Page curve for the de Sitter entropy open for discussion. It could therefore prove useful to employ our method --- that needs neither quantum extremal surfaces nor islands --- to add to this discussion.

\acknowledgments
It is a pleasure to thank Ben Freivogel, Claire Zukowski, Antonio Rotundo and Theodora Nikolakopoulou for discussions during the initial stage of this project. Our work is supported by the Spinoza grant and the Delta ITP consortium, a program of the NWO that is funded by the Dutch Ministry of Education, Culture and Science (OCW). 

\begin{appendices}

\section{Entropy calculations} \label{app:entropy}
In this appendix we provide a detailed computation of the entropies \eqref{eq:extentropy} and \eqref{eq:entbeforePage}/\eqref{eq:entafterPage} using the formula for the geodesic distance $\Delta s$ between two points $s_1, s_2$ in terms of embedding coordinates: 
	\begin{equation}
		-\ell^2 \cosh(\Delta s / \ell) = X^\mu (s_1)X_\mu (s_2)~.
	\end{equation}
The computations are standard \cite{RyuTakayanagi,HeadrickRTlectures}; we added them for completeness. 

\subsection{Extremal \texorpdfstring{AdS$_2$}{AdS2} black hole} \label{sec:appext}
For the Poincar\'e metric, the embedding coordinates are
	\begin{equation}
	\begin{aligned}
		X^0 &= \frac{z}{2} + \frac{1}{2z}(\ell^2 + x^2 - t^2)~, \\
		X^1 &= \frac{\ell^2}{z} t~, \\
		X^2 &= \frac{\ell^2}{z} x ~,\\
		X^3 &= \frac{z}{2} - \frac{1}{2z} (\ell^2 - x^2 + t^2)~,
	\end{aligned}
	\end{equation}
where $X^0$ and $X^1$ are timelike, i.e.\ $-(X^0)^2 - (X^1)^2 + (X^2)^2 + (X^3)^2 = -\ell^2$. The geodesic distance between two points $s_1 = (t_1, z_1, x_1)$ and $s_2 = (t_2, z_2, x_2)$ is then given by 
	\begin{equation}
		\cosh (\Delta s / \ell) =  \frac{1}{2 z_1 z_2} \lp - (t_2 - t_1)^2 + (x_2 - x_1)^2 + z_1^2 + z_2^2 \rp~. 
	\end{equation}
For a fixed time slice, the geodesic distance depends only on $z_1, z_2$ and $\Delta x$. If the endpoints lie at the same radial distance $z_1= z_2 = z$ we find
	\begin{equation}
		\cosh (\Delta s / \ell ) = 1 + \frac{1}{2} \lp \frac{\Delta x}{z }\rp^2 \, \Rightarrow \, \Delta s = 2\ell \asinh \frac{\Delta x}{2z}~.
	\end{equation}
For $z \to \epsilon \ell$ with $\epsilon\to 0$ (i.e.\ points close to the boundary) we can approximate	
	\begin{equation}
		\Delta s = 2\ell \log \frac{\Delta x}{2\ell} \,+ \text{ UV cutoff}~.
	\end{equation}
Mapping the Poincar\'e patch to the cylinder using $x = \ell \varphi$, $z = \frac{\ell^2}{r}$, as in the discussion around \eqref{eq:Poincarecylin}, we find the entropy of an angular interval $\Delta \varphi$ to be
	\begin{equation}
		S = \frac{1}{4G} \lp 2\log{\Delta \varphi} \rp~, 
	\end{equation}
where we used $G^{(3)} = \ell G^{(2)}$ and dropped the UV cutoff. For the angular interval in figure \ref{fig:geod1}, we have $\Delta \varphi = 2\pi \alpha + \frac{2b}{\ell} $. Then we can distinguish two regimes $b \ll \Phi_r$ and $b \gg \Phi_r$, leading to 
	\begin{equation}
		S = \begin{cases}\frac{1}{4G} \left( 2\log{\Phi_r / \ell} \right) & \mbox{if } b \ll \Phi_r \\ \frac{1}{4G} \left( 2\log{2b / \ell} \right) & \mbox{if } b \gg \Phi_r \end{cases} ~,
	\end{equation}
where we used $\Phi_r = 2\pi \ell \alpha$.

\subsection{Finite temperature \texorpdfstring{AdS$_2$}{AdS2} black hole} \label{sec:appfinite}
For the BTZ metric \eqref{eq:BTZ} and \eqref{eq:BTZmetric2} the embedding coordinates are 
	\begin{equation}
		\begin{aligned}
			X^0 &= \frac{\ell}{R} \sqrt{r^2 - R^2} \sinh \frac{Rt}{\ell^2} = \ell \frac{\sinh \frac{\pi}{\beta} (u+v)}{\sinh \frac{\pi}{\beta}(u-v)}~, \\ 
			X^1 &= \frac{r \ell}{R} \cosh \frac{R \varphi}{\ell} = \frac{\cosh \frac{2\pi}{\beta} \ell \varphi}{\tanh \frac{\pi}{\beta} (u-v)}~, \\ 
			X^2 &= \frac{r \ell}{R} \sinh \frac{R \varphi}{\ell} = \frac{\sinh \frac{2\pi}{\beta} \ell \varphi}{\tanh \frac{\pi}{\beta} (u-v)}~, \\
			X^3 &= \frac{\ell}{R} \sqrt{r^2 - R^2} \cosh \frac{Rt}{\ell^2} = \ell \frac{\cosh \frac{\pi}{\beta} (u+v)}{\sinh \frac{\pi}{\beta}(u-v)}~,
		\end{aligned}
	\end{equation}
where again $X^0$ and $X^1$ are timelike directions. For two points $s_1 = (t_1, r_1, \varphi_1)$ and $s_2 = (t_2, r_2, \varphi_2)$ the geodesic distance is
	\begin{equation}
		\cosh (\Delta s / \ell) = \frac{r_1 r_2}{R^2} \cosh \frac{R}{\ell} (\varphi_1 - \varphi_2) - \sqrt{\lp\frac{r_1^2}{R^2} - 1 \rp \lp \frac{r_2^2}{R^2} -1 \rp } \cosh{ \frac{R}{\ell} (t_1-t_2)}~,
	\end{equation}
such that for a fixed time slice and for two points at the same radius $r_1=r_2=r$ one finds
	\begin{equation}
		1 - \cosh (\Delta s /\ell) = \frac{r^2}{R^2} \lp 1 - \cosh{\frac{R \Delta \varphi}{\ell}} \rp \, \Rightarrow \Delta s = 2 \ell \asinh \frac{r}{R} \sinh \frac{R \Delta \varphi}{2\ell}~. 
	\end{equation}	
Now, for $r \to \infty$ we can expand to find
	\begin{equation}
		\Delta s = 2 \ell \log{\sinh{\frac{R\Delta \varphi}{2\ell}}} \, + \, \text{UV cutoff}~,
	\end{equation}
leading to an entropy 
	\begin{equation} \label{eq:finentropyinterval}
	 	S = \frac{1}{4G^{(2)}} \left(2 \log{ \sinh{ \frac{\pi}{\beta} \ell \Delta \varphi}} \right)~,
	\end{equation}
where we used $G^{(3)} = \ell G^{(2)}$, $R = \frac{2\pi \ell^2}{\beta}$ and dropped the UV cutoff.
\end{appendices}

\bibliographystyle{JHEP}
\bibliography{library}

\end{document}